\documentclass[aps,prb,reprint,showpacs,superscriptaddress,preprintnumbers]{revtex4-1}
\usepackage{amsmath,times,bm,graphicx,color,hyperref}
\newcommand\cc{{\rm c.c.}}

\newcommand\dc[1]{\dot {\cal #1}}

\newcommand\hb[1]{\hat {\bm #1}}

\DeclareMathOperator\diag{diag}
\DeclareMathOperator\tr{tr}
\begin{document}
\title{Bulk angular momentum and Hall viscosity in chiral superconductors}
\author{Atsuo Shitade}
\affiliation{Department of Physics, Kyoto University, Kyoto 606-8502, Japan}
\author{Taro Kimura}
\affiliation{Institut de Physique Th\'{e}orique, CEA Saclay, 91191 Gif-sur-Yvette, France}
\affiliation{Mathematical Physics Laboratory, RIKEN Nishina Center, Saitama 351-0198, Japan}
\preprint{IPHT-T14/097, RIKEN-MP-92}
\date{\today}
\begin{abstract}
  We establish the Berry-phase formulas for the angular momentum (AM) and the Hall viscosity (HV)
  to investigate chiral superconductors (SCs) in two and three dimensions.
  The AM is defined by the temporal integral of the antisymmetric momentum current induced by an adiabatic deformation,
  while the HV is defined by the symmetric momentum current induced by the symmetric torsional electric field.
  Without suffering from the system size or geometry,
  we obtain the macroscopic AM $L_z = \hbar m N_0/2$ at zero temperature in full-gap chiral SCs,
  where $m$ is the magnetic quantum number and $N_0$ is the total number of electrons.
  We also find that the HV is equal to half the AM at zero temperature not only in full-gap chiral SCs as is well known but also in nodal ones,
  but its behavior at finite temperature is different in the two cases.
\end{abstract}
\maketitle
\section{Introduction} \label{sec:intro}
Chiral superfluids and superconductors (SCs) are exotic states
whose time-reversal symmetry is spontaneously broken and Cooper pairs carry nonzero angular momentum (AM). 
One well-known example is $^3$He-A whose pairing symmetry is $p_x + i p_y$~\cite{RevModPhys.47.331}.
Among electron systems, there are a few candidates for chiral SCs such as
Sr$_2$RuO$_4$ with $p_x + i p_y$~\cite{RevModPhys.75.657,JPSJ.81.011009}
and URu$_2$Si$_2$ with $d_{zx} + i d_{yz}$~\cite{PhysRevLett.99.116402,PhysRevLett.100.017004,1367-2630-11-5-055061,JPSJ.81.023704}.
Recently the $d_{x^2 - y^2} + i d_{xy}$ pairing symmetry was theoretically proposed in SrPtAs~\cite{PhysRevB.86.100507,PhysRevB.87.180503,PhysRevB.89.020509}.

There is a long-standing problem on the AM in chiral $\ell$-wave SCs, the so-called intrinsic AM paradox.
This paradox is summarized as $L_z = \hbar m N_0/2 \times (\Delta_0/E_{\rm F})^{\gamma}$,
where $|m| \leq \ell$, $N_0$, $\Delta_0$, and $E_{\rm F}$ are
the magnetic quantum number, the total number of electrons, the gap strength, and the Fermi energy, respectively.
$\gamma = 0$~\cite{Ishikawa01061977,Ishikawa01041980,PhysRevB.21.980,volovik1995,JPSJ.67.216,Goryo1998549,%
PhysRevB.69.184511,PhysRevB.84.214509,PhysRevB.85.100506} is the most natural if all electrons form Cooper pairs with the AM $\ell_z = \hbar m$.
On the other hand, $\gamma = 1$~\cite{PhysRev.123.1911,RevModPhys.47.331} is intuitively plausible if a few electrons near the Fermi surface form Cooper pairs.
$\gamma = 2$ was also proposed by using the Ginzburg-Landau theory~\cite{volovik1975,cross1975}.
Recent microscopic studies have settled this paradox, at least theoretically,
to $\gamma = 0$~\cite{JPSJ.67.216,PhysRevB.69.184511,PhysRevB.84.214509,PhysRevB.85.100506}
but have been employed only in finite systems such as a cylinder~\cite{JPSJ.67.216} and a disk~\cite{PhysRevB.69.184511,PhysRevB.84.214509,PhysRevB.85.100506}.

Generally, the physical quantities involving the position operator are ill defined in periodic systems.
The most famous examples are the charge polarization and the orbital magnetization.
The former is classically coupled to an electric field but is quantum-mechanically defined by
the temporal integral of the charge current under an adiabatic deformation~\cite{PhysRevB.47.1651,PhysRevB.49.14202,PhysRevB.88.155121,JPSJ.83.033708}.
On the other hand, the latter is coupled to a magnetic field
and can be defined by the magnetic field derivative of the free energy~\cite{PhysRevLett.99.197202,PhysRevB.84.205137,PhysRevB.86.214415}.
Consequently, the charge polarization is associated with the Berry connection in the Bloch basis,
while the orbital magnetization with the Berry curvature and the magnetic moment.
These formalisms were extended to heat analogs such as the heat polarization~\cite{JPSJ.83.033708} and magnetization~\cite{1310.8043}.
However, the AM has not been formulated yet.

Another interesting clue to the AM is the Hall viscosity (HV),
which has been intensively discussed in the context of the quantum Hall effect~\cite{PhysRevLett.75.697,avron1998}.
Similar to the Hall conductivity, the HV is nonzero only when the time-reversal symmetry is broken.
The important relation $\eta_{\rm H} = \hbar N_0 {\bar s}/2$ holds
in general gapped systems at zero temperature~\cite{PhysRevB.79.045308,PhysRevB.84.085316,PhysRevB.89.174507},
in which the orbital spin ${\bar s}$ is equal to $\ell/2$ in chiral $\ell$-wave SCs.
In addition, it was also related to the momentum-dependent Hall conductivity~\cite{PhysRevLett.108.066805,PhysRevB.86.245309}.
The torsional Chern-Simons term was proposed to describe the quantum HV
in the two-dimensional massive Dirac system~\cite{PhysRevLett.107.075502,Hidaka01012013,PhysRevD.88.025040}.

In this paper, we derive the Berry-phase formulas for the AM and the HV to apply to chiral SCs in two and three dimensions.
The antisymmetric and symmetric components of a torsional electric field describe an angular velocity and a strain-rate tensor, respectively.
Since the AM is conjugate to the former, it can be formulated in the same way as in the charge polarization,
namely, by the temporal integral of the antisymmetric momentum current induced by an adiabatic deformation.
Viscosity is defined by the symmetric momentum current, i.e., the stress tensor, induced by the latter.
In contrast to the previous works regarding the intrinsic AM paradox, we obtain $L_z = \hbar m N_0/2$ without suffering from the finite-size effects.
We also investigate the temperature dependence of the HV for two-dimensional gapped chiral SCs and three-dimensional nodal ones.

Hereafter we assign the Latin ($a, b, \dots = {\hat 0}, {\hat 1}, \dots, {\hat d}$) and Greek ($\mu, \nu, \dots = 0, 1, \dots, d$) alphabets
to locally flat and global coordinates, respectively.
We follow the Einstein convention, which implies summation over the spacetime dimension $D = d + 1$ when an index appears twice in a single term.
The Minkowski metric is taken as $\eta_{ab} = \diag (-1, +1, \dots, +1)$.
The Planck constant and the charge are denoted by $\hbar$ and $q$, while the speed of light and the Boltzmann constant are put to $c = k_{\rm B} = 1$.
The upper or lower signs in equations correspond to boson or fermion.

\section{Cartan Formalism} \label{sec:cartan}
To begin with, we examine an angular velocity from the gauge-theoretical viewpoint.
Now that the system is rotated, we have to deal with a theory in a curved spacetime.
Here we use the Cartan formalism, which consists of two gauge potentials
such as a vielbein and a spin connection~\cite{9789812791719}.
A vielbein $h^a_{\phantom{a} \mu}$ is a gauge potential corresponding to local spacetime translations,
while a spin connection $\omega^{ab}_{\phantom{ab} \mu}$ is that corresponding to local Lorentz transformations.
In these gauge potentials as well as a vector potential $A_{\mu}$, the partial derivative is replaced by the covariant derivative
$\partial_a \to D_a \equiv h_a^{\phantom{a} \mu} (\partial_{\mu} - i q A_{\mu}/\hbar - i \omega^{ab}_{\phantom{\mu}} S_{ab}/2 \hbar)$.
Here $S_{ab}$ is the generator of local Lorentz transformations.
The spatial component of a vielbein is related to a displacement vector~\cite{PhysRevLett.107.075502,Hidaka01012013,PhysRevD.88.025040}
as $h^{\hat k}_{\phantom{\hat k} i} = \delta^{\hat k}_{\phantom{\hat k} i} + \partial_i u^{\hat k}$.
Moreover, $h^{\hat k}_{\phantom{\hat k} 0}$ describes rotation as illustrated below.

For simplicity, let us consider a Dirac fermion in a curved spacetime.
The Dirac Lagrangian density is given by
\begin{equation}
  h {\cal L}
  = h {\bar \psi} (\hbar \gamma^a D_a - m) \psi, \label{eq:dirac1}
\end{equation}
in which $h \equiv \det h^a_{\phantom{a} \mu}$ is the determinant of a vielbein,
${\bar \psi} \equiv i^{-1} \psi^{\dag} \gamma^{\hat 0}$ is the Dirac conjugate,
and the $\gamma$ matrices satisfy the Clifford algebra $\{\gamma^a, \gamma^b\} = 2 \eta^{ab}$.
When we introduce the nonzero off-diagonal component $h^{\hat k}_{\phantom{\hat k} 0} = \phi_{\rm r}^{\hat k}$ in addition to the identity,
the square line element is given by
\begin{equation}
  d s^2
  = -d t^2 + (d {\vec x} + {\vec \phi}_{\rm r} d t)^2, \label{eq:line}
\end{equation}
and the Dirac Lagrangian is reduced to
\begin{equation}
  h {\cal L}
  = \psi^{\dag} [i \hbar D_0 + i \hbar {\vec \alpha} \cdot {\vec D} - m \beta - i \hbar {\vec \phi}_{\rm r} \cdot {\vec D}] \psi. \label{eq:dirac2}
\end{equation}
Here ${\vec \alpha}$ and $\beta$ are the Dirac matrices.
If we put ${\vec \phi}_{\rm r} = {\vec \Omega} \times {\vec x}$, ${\vec \Omega}$ is assigned to an angular velocity,
and the fourth term represents the coupling between an angular velocity and the AM.
Such a discussion relies on the translational symmetry and gauge principle of gravity and hence is not restricted to relativistic systems.

Since a vielbein is a gauge potential, it induces a field strength called torsion.
Here a spin connection is neglected, and hence torsion is written by
\begin{subequations} \begin{align}
  T^{\hat l}_{\phantom{\hat l} j0}
  = & \partial_j h^{\hat l}_{\phantom{\hat l}0} - \partial_0 h^{\hat l}_{\phantom{\hat l} j}, \label{eq:torsion1a} \\
  T^{\hat l}_{\phantom{\hat l} ij}
  = & \partial_i h^{\hat l}_{\phantom{\hat l} j} - \partial_j h^{\hat l}_{\phantom{\hat l} i}. \label{eq:torsion1b}
\end{align} \label{eq:torsion1} \end{subequations}
The former is ``electric.''
The first term describes an angular velocity if ${\hat l}$ and $j$ are antisymmetric,
while the second term describes a strain-rate tensor if symmetric.
On the other hand, the latter is ``magnetic'' and characterizes edge and screw dislocations in crystals.
Especially its flux is identified as the Burgers vector.

It is a natural question why rotation is described by a vielbein, i.e., a gauge potential corresponding to space translations.
In fact, global space translations do not include rotations.
However, a gauge potential is associated with a local symmetry, and local space translations do include rotations.
Note that an angular velocity is coupled not only to the AM but to the spin, which is implemented by a spin connection.
Such spin responses are out of scope here.

\section{Angular Momentum} \label{sec:am}
Since an angular velocity is ``electric,'' the AM is similar to the charge polarization rather than to the orbital magnetization.
Below we define the momentum polarization by the temporal integral of the nonsymmetric momentum current induced by an adiabatic deformation.
The momentum current is conjugate to a ``vector potential'' $h^{\hat k}_{\phantom{\hat k} i}$
and is given by the product of the momentum and the velocity.
In the Wigner representation of the Keldysh formalism, it is written by
\begin{equation}
  T_{\hat k}^{\phantom{\hat k} {\hat \imath}}
  = \pm \frac{i \hbar}{2} \int \frac{d^D \pi}{(2 \pi \hbar)^D}
  \tr [(-\partial_{\pi_{\hat \imath}} {\hb G}^{-1}) \star {\hb G} \star \pi_{\hat k}]^< + \cc \label{eq:momcurr}
\end{equation}
Here ${\hb G}$ is the Keldysh Green's function,
$(-\partial_{\pi_{\hat \imath}} {\hb G}^{-1})$ is the renormalized velocity to satisfy the local conservation law,
and $\pi_{\hat k}$ is the Wigner representation of the covariant derivative corresponding to the momentum.
Note that this momentum current is Hermitian but not symmetric over ${\hat k}$ and ${\hat \imath}$.

An adiabatic deformation is implemented by the gradient expansion up to the first order~\cite{JPSJ.83.033708}.
Since both the Keldysh Green's function ${\hb G}$ and the renormalized velocity $(-\partial_{\pi_{\hat \imath}} {\hb G}^{-1})$ are perturbed,
the gradient expansion of the momentum current Eq.~\eqref{eq:momcurr} is given by
\begin{align}
  T_{\hat k}^{\phantom{\hat k} {\hat \imath}}
  = & \pm \frac{i \hbar^2}{4} \int \frac{d^D \pi}{(2 \pi \hbar)^D} \pi_{\hat k}
  \tr [\partial_{\pi_{\hat \imath}} {\hat \Sigma}_1 {\hat G}_0 - \partial_{\pi_{\hat \imath}} {\hat G}_0^{-1} {\hat G}_1]^< \notag \\
  & + \cc, \label{eq:mp1}
\end{align}
in which the gradient expansion of the Keldysh Green's function ${\hat G}_1$ is given by
\begin{equation}
  {\hat G}_1
  = {\hat G}_0 {\hat \Sigma}_1 {\hat G}_0
  + i [{\hat G}_0 \partial_{X^0} {\hat G}_0^{-1} {\hat G}_0 \partial_{\pi_{\hat 0}} {\hat G}_0^{-1} {\hat G}_0 - (X^0 \leftrightarrow \pi_{\hat 0})], \label{eq:gd}
\end{equation}
and the corresponding self-energy ${\hat \Sigma}_1$ is determined self-consistently.
Now the momentum current is transformed into
\begin{widetext}
\begin{align}
  T_{\hat k}^{\phantom{\hat k} {\hat \imath}}
  = & \pm \frac{i \hbar^2}{4} \int \frac{d^D \pi}{(2 \pi \hbar)^D} \pi_{\hat k}
  \tr [\partial_{\pi_{\hat \imath}} {\hat \Sigma}_1 {\hat G}_0 - \partial_{\pi_{\hat \imath}} {\hat G}_0^{-1} {\hat G}_0 {\hat \Sigma}_1 {\hat G}_0]^<
  + \cc \notag \\
  & \pm \frac{\hbar^2}{4} \int \frac{d^D \pi}{(2 \pi \hbar)^D} \pi_{\hat k}
  \tr [\partial_{\pi_{\hat \imath}} {\hat G}_0^{-1} {\hat G}_0 \partial_{X^0} {\hat G}_0^{-1} {\hat G}_0 \partial_{\pi_{\hat 0}} {\hat G}_0^{-1} {\hat G}_0
  - (X^0 \leftrightarrow \pi_{\hat 0})]^< + \cc \label{eq:mp2}
\end{align}
By extracting the lesser component and employing the temporal integral,
we obtain the change in the momentum polarization, but not the momentum polarization itself,
\begin{subequations} \begin{align}
  \Delta P_{\hat k}^{\phantom{\hat k} {\hat \imath}}
  \equiv & \int d X^0 T_{\hat k}^{\phantom{\hat k} {\hat \imath}} \notag \\
  = & \frac{\hbar^2}{6} \epsilon_{ABC} \int \frac{d^D \pi}{(2 \pi \hbar)^D} \int d X^0 f(-\pi_{\hat 0}) \pi_{\hat k}
  \tr [G_0^{\rm R} \partial_A G_0^{{\rm R} -1} G_0^{\rm R} \partial_B G_0^{{\rm R} -1} G_0^{\rm R} \partial_C G_0^{{\rm R} -1}] + \cc \label{eq:mp3a} \\
  & + \frac{\hbar^2}{4} \int \frac{d^D \pi}{(2 \pi \hbar)^D} \int d X^0 f^{\prime}(-\pi_{\hat 0}) \pi_{\hat k}
  \tr [(G_0^{\rm R} - G_0^{\rm A}) \partial_{\pi_{\hat \imath}} (G_0^{{\rm R} -1} + G_0^{{\rm A} -1}) G_0^{\rm R} \partial_{X^0} G_0^{{\rm R} -1}]
  + \cc \label{eq:mp3b} \\
  & \mp \frac{i \hbar^2}{4} \delta_{\hat k}^{\phantom{\hat k} {\hat \imath}} \int \frac{d^D \pi}{(2 \pi \hbar)^D} \int d X^0
  \tr [{\hat \Sigma}_1 {\hat G}_0]^< + \cc \label{eq:mp3c} \\
  & \pm \frac{i \hbar^2}{4} \int \frac{d^D \pi}{(2 \pi \hbar)^D} \int d X^0 f^{\prime}(-\pi_{\hat 0}) \pi_{\hat k}
  \tr [\Sigma_1^{< (1)} (G_0^{\rm R} - G_0^{\rm A}) \partial_{\pi_{\hat \imath}} G_0^{{\rm R} -1} G_0^{\rm R}] + \cc, \label{eq:mp3d}
\end{align} \label{eq:mp3} \end{subequations}
\end{widetext}
where $\epsilon_{ABC}$ in Eq.~\eqref{eq:mp3a} is the antisymmetric tensor with $\epsilon_{\pi_{\hat 0} \pi_{\hat \imath} X^0} = 1$.
$f(\omega) = (e^{\beta \omega} \mp 1)^{-1}$ is the distribution function for boson or fermion.

Below we focus on the clean and noninteracting limit ${\hb \Sigma} = 0$ to derive the Berry-phase formula in the Bloch basis.
The retarded Green's function is given by $G_0^{\rm R} = [(-\pi_{\hat 0}) - {\cal H}(X^0) + \mu + i \eta]^{-1}$ with $\eta \to +0$,
leading to $\partial_{\pi_{\hat 0}} G_0^{{\rm R} -1} = -1$, $\partial_{\pi_{\hat \imath}} G_0^{{\rm R} -1} = -v^{\hat \imath}$,
and $\partial_{X^0} G_0^{{\rm R} -1} = - {\dc H}$.
In an adiabatic deformation, the trace can be expanded with respect to the eigenstates satisfying
${\cal H}(X^0) | u_{n {\vec \pi} X^0} \rangle = \epsilon_{n {\vec \pi} X^0} | u_{n {\vec \pi} X^0} \rangle$.
We evaluate the integral over $(-\pi_{\hat 0})$ by the residue theorem to obtain
\begin{equation}
  \Delta P_{\hat k}^{\phantom{\hat k} {\hat \imath}}
  = -\sum_n \int \frac{d^d \pi}{(2 \pi \hbar)^d} \int d X^0 \pi_{\hat k}
  \Omega^{\hat \imath}_{n {\vec \pi} X^0} f_{n {\vec \pi} X^0}, \label{eq:mp4}
\end{equation}
with $f_{n {\vec \pi} X^0} \equiv f(\epsilon_{n {\vec \pi} X^0} - \mu)$ and the Berry curvature in the $(\pi_{\hat \imath}/\hbar, X^0)$-space being defined by
\begin{equation}
  \Omega_{n {\vec \pi} X^0}^{\hat \imath}
  \equiv i \hbar \langle \partial_{\pi_{\hat \imath}} u_{n {\vec \pi} X^0} | \partial_{X^0} u_{n {\vec \pi} X^0} \rangle
  - (X^0 \leftrightarrow \pi_{\hat \imath}). \label{eq:berrycurv-t}
\end{equation}
Similar to the charge polarization, Eq.~\eqref{eq:mp4} depends on the choice of an adiabatic deformation at finite temperature and hence is not well defined.
At zero temperature in a gapped fermion system, its integrand becomes the total derivative with respect to $X^0$ when ${\hat k} \not= {\hat \imath}$,
and the momentum polarization itself is given by
\begin{equation}
  P_{\hat k}^{\phantom{\hat k} {\hat \imath}}
  = \sum_n^{\rm occ} \int \frac{d^d \pi}{(2 \pi \hbar)^d} \pi_{\hat k} A_{n {\vec \pi}}^{\hat \imath}, \label{eq:mp5}
\end{equation}
where we introduce the Berry connection,
\begin{equation}
  A_{n {\vec \pi}}^{\hat \imath}
  \equiv i \hbar \langle u_{n {\vec \pi}} | \partial_{\pi_{\hat \imath}} u_{n {\vec \pi}} \rangle. \label{eq:berrycon}
\end{equation}
Indeed this expression is quite similar to that for the charge polarization, which is given by the integral of the Berry connection itself.
Note that at the initial time, the reference system has time-reversal symmetry, and the momentum polarization should be zero.
After all, the AM is obtained by the antisymmetric part of the momentum polarization,
\begin{equation}
  L_{\hat k}
  \equiv \epsilon_{{\hat \imath} {\hat \jmath} {\hat k}} P^{{\hat \jmath} {\hat \imath}}
  = \sum_n^{\rm occ} \int \frac{d^d \pi}{(2 \pi \hbar)^d} \epsilon_{{\hat \imath} {\hat \jmath} {\hat k}}
  A^{\hat \imath}_{n {\vec \pi}} \pi^{\hat \jmath}. \label{eq:am}
\end{equation}
Since the Berry connection is regarded as the expectation value of the position operator in the Wannier basis,
this Berry-phase formula really indicates ${\vec \ell} = {\vec x} \times {\vec p}$ in the momentum space.

In the above derivation, it is not obvious why the system should be gapped.
Wave functions have the phase degree of freedom; namely,
physical quantities should be invariant under a unitary transformation $| u^{\prime}_{n {\vec \pi}} \rangle = e^{-i \theta_{n {\vec \pi}}} | u_{n {\vec \pi}} \rangle$.
Correspondingly, the Berry connection Eq.~\eqref{eq:berrycon} is transformed as
${\vec A}_{n {\vec \pi}}^{\prime} = {\vec A}_{n {\vec \pi}} + \hbar {\vec \partial}_{\pi} \theta_{n {\vec \pi}}$,
and the integrand in Eq.~\eqref{eq:am} is transformed as
${\vec A}_{n {\vec \pi}}^{\prime} \times {\vec \pi}
= {\vec A}_{n {\vec \pi}} \times {\vec \pi} + \hbar {\vec \partial}_{\pi} \times (\theta_{n {\vec \pi}} {\vec \pi})$.
In a gapless system where the momentum space is restricted, the AM is not invariant.
On the other hand, in a gapped system, the AM is found to be invariant by using the Stokes theorem and the single-valued property of wave functions.

\section{Hall Viscosity} \label{sec:eta}
Here we begin with a brief introduction of elasticity and viscosity.
These are the mechanical properties of a system and are characterized by
\begin{equation}
  T^{(ki)}
  = \lambda^{(ki)(lj)} u_{(lj)} + \eta^{(ki)(lj)} {\dot u}_{(lj)}. \label{eq:visco}
\end{equation}
Here $T^{(ki)}$, $u_{(lj)}$, and ${\dot u}_{(lj)}$ are stress, strain, and strain-rate tensors, respectively, and round brackets indicate the symmetry over their indexes.
The linear coefficients $\lambda^{(ki)(lj)}$ and $\eta^{(ki)(lj)}$ are dubbed the elastic modulus and the viscosity, although their sign convention is not fixed.
We can decompose the viscosity into the symmetric and antisymmetric parts: $\eta^{(ki)(lj)} = \eta^{(ki)(lj)}_{\rm S} + \eta^{(ki)(lj)}_{\rm A}$,
where $\eta^{(ki)(lj)}_{\rm S} = \eta^{(lj)(ki)}_{\rm S}$ and $\eta^{(ki)(lj)}_{\rm A} = -\eta^{(lj)(ki)}_{\rm A}$.
The latter is dubbed the HV and is formulated below.

As discussed above, a strain-rate tensor is the symmetric torsional electric field, and the stress tensor is the symmetric part of the momentum current.
Therefore the viscosity can be formulated by the perturbation theory of the momentum current Eq.~\eqref{eq:momcurr} with respect to torsion~\cite{1310.8043}.
Here we define the nonsymmetric viscosity by
\begin{align}
  \eta_{{\hat k} \phantom{\hat \imath} {\hat l}}^{\phantom{\hat k} {\hat \imath} \phantom{\hat l} {\hat \jmath}}
  \equiv & \frac{\partial T_{\hat k}^{\phantom{\hat k} {\hat \imath}}}{\partial (-T^{\hat l}_{\phantom{\hat l} {\hat \jmath} {\hat 0}})} \notag \\
  = & \mp \frac{i \hbar^2}{2} \int \frac{d^D \pi}{(2 \pi \hbar)^D} \pi_{\hat k}
  \tr [\partial_{\pi_{\hat \imath}} {\hat \Sigma}_{T^{\hat l}_{\phantom{\hat l} {\hat \jmath} {\hat 0}}} {\hat G}_0
  - \partial_{\pi_{\hat \imath}} {\hat G}_0^{-1} {\hat G}_{T^{\hat l}_{\phantom{\hat l} {\hat \jmath} {\hat 0}}}]^< \notag \\
  & + \cc \label{eq:eta1}
\end{align}
Note that the negative sign in the definition is necessary
because $\partial_0 h^{\hat l}_{\phantom{\hat l} j}$ in Eq.~\eqref{eq:torsion1a} gives rise to a strain-rate tensor.
The Keldysh Green's function in the presence of torsion is given by
\begin{align}
  {\hat G}_{T^a_{\phantom{a} cd}}
  = & {\hat G}_0 {\hat \Sigma}_{T^a_{\phantom{a} cd}} {\hat G}_0 \notag \\
  & - \pi_a [{\hat G}_0 \partial_{\pi_c} {\hat G}_0^{-1} {\hat G}_0 \partial_{\pi_d} {\hat G}_0^{-1} {\hat G}_0 - (c \leftrightarrow d)]/2 i, \label{eq:gtem}
\end{align}
\begin{widetext}
and the corresponding self-energy $\Sigma_{T^a_{\phantom{a} cd}}$ is determined self-consistently.
Equation~\eqref{eq:eta1} is rewritten by
\begin{align}
  \eta_{{\hat k} \phantom{\hat \imath} {\hat l}}^{\phantom{\hat k} {\hat \imath} \phantom{\hat l} {\hat \jmath}}
  = & \mp \frac{i \hbar^2}{2} \int \frac{d^D \pi}{(2 \pi \hbar)^D} \pi_{\hat k}
  \tr [\partial_{\pi_{\hat \imath}} {\hat \Sigma}_{T^{\hat l}_{\phantom{\hat l} {\hat \jmath} {\hat 0}}} {\hat G}_0
  - \partial_{\pi_{\hat \imath}} {\hat G}_0^{-1} {\hat G}_0 {\hat \Sigma}_{T^{\hat l}_{\phantom{\hat l} {\hat \jmath} {\hat 0}}} {\hat G}_0]^< + \cc \notag \\
  & \mp \frac{\hbar^2}{4} \int \frac{d^D \pi}{(2 \pi \hbar)^D} \pi_{\hat k} \pi_{\hat l}
  \tr [\partial_{\pi_{\hat \imath}} {\hat G}_0^{-1}
  {\hat G}_0 \partial_{\pi_{\hat \jmath}} {\hat G}_0^{-1} {\hat G}_0 \partial_{\pi_{\hat 0}} {\hat G}_0^{-1} {\hat G}_0
  - (j \leftrightarrow 0)]^< + \cc, \label{eq:eta2}
\end{align}
and then
\begin{subequations} \begin{align}
  \eta_{{\hat k} \phantom{\hat \imath} {\hat l}}^{\phantom{\hat k} {\hat \imath} \phantom{\hat l} {\hat \jmath}}
  = & -\frac{\hbar^2}{6} \epsilon^{{\hat \imath} {\hat \jmath} {\hat k}} \epsilon_{abc {\hat k}}
  \int \frac{d^D \pi}{(2 \pi \hbar)^D} f(-\pi_{\hat 0}) \pi_{\hat k} \pi_{\hat l}
  \tr [G_0^{\rm R} \partial_{\pi_a} G_0^{{\rm R} -1} G_0^{\rm R} \partial_{\pi_b} G_0^{{\rm R} -1}
  G_0^{\rm R} \partial_{\pi_c} G_0^{{\rm R} -1}] + \cc \label{eq:eta3a} \\
  & - \frac{\hbar^2}{4} \int \frac{d^D \pi}{(2 \pi \hbar)^D} f^{\prime}(-\pi_{\hat 0}) \pi_{\hat k} \pi_{\hat l}
  \tr [(G_0^{\rm R} - G_0^{\rm A}) \partial_{\pi_{\hat \imath}} (G_0^{{\rm R} -1} + G_0^{{\rm A} -1})
  G_0^{\rm R} \partial_{\pi_{\hat \jmath}} G_0^{{\rm R} -1}] + \cc \label{eq:eta3b} \\
  & \pm \frac{i \hbar^2}{2} \delta_{\hat k}^{\phantom{\hat k} {\hat \imath}} \int \frac{d^D \pi}{(2 \pi \hbar)^D}
  \tr [{\hat \Sigma}_{T^{\hat l}_{\phantom{\hat l} {\hat \jmath} {\hat 0}}} {\hat G}_0]^< + \cc \label{eq:eta3c} \\
  & \mp \frac{i \hbar^2}{2} \int \frac{d^D \pi}{(2 \pi \hbar)^D} f^{\prime}(-\pi_{\hat 0}) \pi_{\hat k}
  \tr [\Sigma_{T^{\hat l}_{\phantom{\hat l} {\hat \jmath} {\hat 0}}}^{< (1)} (G_0^{\rm R} - G_0^{\rm A})
  \partial_{\pi_{\hat \imath}} G_0^{{\rm R} -1} G_0^{\rm R}] + \cc \label{eq:eta3d}
\end{align} \label{eq:eta3} \end{subequations}
\end{widetext}

Again we focus on the clean and noninteracting limit ${\hb \Sigma} = 0$.
The retarded Green's function is given by $G_0^{\rm R} = [(-\pi_{\hat 0}) - {\cal H} + \mu + i \eta]^{-1}$.
We expand the trace with respect to the Bloch basis and evaluate the integral over $(-\pi_{\hat 0})$ by the residue theorem.
As a result, we obtain the Berry-phase formula for the nonsymmetric HV,
\begin{equation}
  \eta_{{\hat k} \phantom{\hat \imath} {\hat l}}^{\phantom{\hat k} {\hat \imath} \phantom{\hat l} {\hat \jmath}}
  = \frac{1}{\hbar} \epsilon^{{\hat \imath} {\hat \jmath} {\hat m}} \sum_n \int \frac{d^d \pi}{(2 \pi \hbar)^d} \pi_{\hat k} \pi_{\hat l}
  \Omega_{n {\vec \pi} {\hat m}} f_{n {\vec \pi}}, \label{eq:eta4}
\end{equation}
with $f_{n {\vec \pi}} \equiv f(\epsilon_{n {\vec \pi}} - \mu)$ and the Berry curvature being defined by
\begin{equation}
  \Omega_{n {\vec \pi} {\hat k}}
  = i \hbar^2 \epsilon_{{\hat \imath} {\hat \jmath} {\hat k}} \langle \partial_{\pi_{\hat \imath}} u_{n {\vec \pi}} |
  \partial_{\pi_{\hat \jmath}} u_{n {\vec \pi}} \rangle. \label{eq:berrycurv}
\end{equation}
The proper HV should be symmetric as shown in Eq.~\eqref{eq:visco}.
In the conventional approach involving a metric,
the stress tensor is defined by $T^{\mu \nu} \equiv 2 \delta S/\delta g_{\mu \nu}$.
This is manifestly symmetric because a metric is symmetric.
By using $g_{\mu \nu} = \eta_{ab} h^a_{\phantom{a} \mu} h^b_{\phantom{b} \nu}$,
it can be related to the momentum current $T_a^{\phantom{a} \mu} \equiv \delta S/\delta h^a_{\phantom{a} \mu}$
by $T^{\mu \nu} = (h^{a \mu} T_a^{\phantom{a} \nu} + h^{a \nu} T_a^{\phantom{a} \mu})/2$.
Therefore the symmetric HV is given by
\begin{equation}
  \eta^{({\hat k} {\hat \imath}) ({\hat l} {\hat \jmath})}
  \equiv (\eta^{{\hat k} {\hat \imath} {\hat l} {\hat \jmath}} + \eta^{{\hat \imath} {\hat k} {\hat l} {\hat \jmath}}
  + \eta^{{\hat k} {\hat \imath} {\hat \jmath} {\hat l}} + \eta^{{\hat \imath} {\hat k} {\hat \jmath} {\hat l}})/4. \label{eq:eta5}
\end{equation}
Although a strain-rate tensor is described by torsion,
Eq.~\eqref{eq:eta5} is different from the torsional HV discussed in Refs.~\onlinecite{PhysRevLett.107.075502,Hidaka01012013,PhysRevD.88.025040}.
In two dimensions, only three components may be nonzero,
\begin{subequations} \begin{align}
  \eta^{(xx)(xy)}
  = & \frac{1}{2 \hbar} \sum_n \int \frac{d^2 \pi}{(2 \pi \hbar)^2} \pi^{x 2} \Omega_{n {\vec \pi} z} f_{n {\vec \pi}}, \label{eq:eta2da} \\
  \eta^{(xx)(yy)}
  = & \frac{1}{\hbar} \sum_n \int \frac{d^2 \pi}{(2 \pi \hbar)^2} \pi^x \pi^y \Omega_{n {\vec \pi} z} f_{n {\vec \pi}}, \label{eq:eta2db} \\
  \eta^{(xy)(yy)}
  = & \frac{1}{2 \hbar} \sum_n \int \frac{d^2 \pi}{(2 \pi \hbar)^2} \pi^{y 2} \Omega_{n {\vec \pi} z} f_{n {\vec \pi}}. \label{eq:eta2dc}
\end{align} \label{eq:eta2d} \end{subequations}
Furthermore, if a system is rotationally invariant, there is only one nonzero component $\eta_{\rm H} = \eta^{(xx)(xy)} = \eta^{(xy)(yy)}$,
\begin{equation}
  \eta_{\rm H}
  = \frac{1}{4 \hbar} \sum_n \int \frac{d^2 \pi}{(2 \pi \hbar)^2} {\vec \pi}^2 \Omega_{n {\vec \pi} z} f_{n {\vec \pi}}. \label{eq:etah}
\end{equation}
In three dimensions, all the components of the HV should vanish if a system is rotationally invariant.
On the other hand, if a system is axially invariant along the $z$ axis, two components are independent,
\begin{subequations} \begin{align}
  \eta^{(xx)(xy)}
  = & \frac{1}{2 \hbar} \sum_n \int \frac{d^3 \pi}{(2 \pi \hbar)^3} \pi^{x 2} \Omega_{n {\vec \pi} z} f_{n {\vec \pi}}, \label{eq:eta3da} \\
  \eta^{(zx)(yz)}
  = & \frac{1}{4 \hbar} \sum_n \int \frac{d^3 \pi}{(2 \pi \hbar)^3} \notag \\
  & \times (\pi^{z 2} \Omega_{n {\vec \pi} z} - \pi^z \pi^x \Omega_{n {\vec \pi} x} - \pi^z \pi^y \Omega_{n {\vec \pi} y}) f_{n {\vec \pi}}. \label{eq:eta3db}
\end{align} \label{eq:eta3d} \end{subequations}
As in the case of the momentum polarization Eq.~\eqref{eq:mp5},
these expressions are quite analogous to that for the Hall conductivity, corresponding to the charge transport.
The integrand just differs in the factor of ${\vec \pi}$.

Finally, let us comment on the interacting cases.
In the conventional metric approach, the effects of interactions are fully taken into account by using the many-body ground-state wave function.
On the other hand, in our approach,  such effects are compiled into the self-energy
and can be taken into account by using the Feynman diagrams or the dynamical mean-field theory.
Therefore these two approaches are complementary and coincide in the noninteracting limit,
which reminds us that the charge polarization can be calculated
by averaging over boundary conditions~\cite{PhysRevB.49.14202} or by using the Green's function formula~\cite{PhysRevB.88.155121,JPSJ.83.033708}.

\section{Applications to Chiral Superconductors} \label{sec:chiral}
Now we apply our results to chiral SCs.
For simplicity, we restrict ourselves to the single-band model with a singlet or unitary triplet pairing described by
\begin{equation}
  H - \mu N
  = \sum_{\vec k}
  \begin{bmatrix}
    c^{\dag}_{{\vec k} \uparrow} & c_{-{\vec k} \downarrow}
  \end{bmatrix}
  \begin{bmatrix}
    \xi_{\vec k} & \Delta_{\vec k} \\
    \Delta^{\ast}_{\vec k} & -\xi_{\vec k}
  \end{bmatrix}
  \begin{bmatrix}
    c_{{\vec k} \uparrow} \\
    c_{-{\vec k} \downarrow}
  \end{bmatrix}. \label{eq:bcs}
\end{equation}
in which $\xi_{\vec k} \equiv \epsilon_{\vec k} - \mu$ is even,
and the gap $\Delta_{\vec k}$ is even or odd for singlet or triplet with ${\vec d}_{\vec k} \parallel {\vec z}$, respectively.
This Hamiltonian has the positive and negative dispersions $\pm E_{\vec k} = \pm \sqrt{\xi_{\vec k}^2 + |\Delta_{\vec k}|^2}$,
whose wave functions are given by
\begin{equation}
  \begin{bmatrix}
    | u_{{\vec k} +} \rangle & | u_{{\vec k} -} \rangle
  \end{bmatrix}
  =
  \begin{bmatrix}
    u_{\vec k} & -v_{\vec k} \\
    v_{\vec k}^{\ast} & u_{\vec k}
  \end{bmatrix}. \label{eq:bdg}
\end{equation}
Here $u_{\vec k} = \sqrt{(1 + \xi_{\vec k}/E_{\vec k})/2}$ and $v_{\vec k} = \sqrt{(1 - \xi_{\vec k}/E_{\vec k})/2} \Delta_{\vec k}/|\Delta_{\vec k}|$
satisfy the normalization condition $u_{\vec k}^2 + |v_{\vec k}|^2 = 1$.
The temperature dependence of the gap strength is obtained by solving the gap equation.
Now that we are not interested in the competition between several phases but in the AM and the HV for chiral SCs,
we assume the attractive interaction favoring a chiral pairing symmetry $\Delta_{\vec k} = \Delta w_{\vec k}$
with $w_{\vec k}$ being normalized by $\int d \Omega_{\vec k} |w_{\vec k}|^2/4 \pi = 1$ in three dimensions.
By using the standard approximations, the gap equation is given by
\begin{equation}
  \ln |\Delta_0/\Delta|
  = \int_0^{\infty} d \xi_{\vec k} \int \frac{d \Omega_{\vec k}}{4 \pi} 2 |w_{\vec k}|^2 f(E_{\vec k})/E_{\vec k}. \label{eq:gap}
\end{equation}

Next, we calculate the Berry connection and curvature.
The Berry connections for the positive- and negative-energy states have opposite signs; namely,
$A_{+ {\vec k}}^{\hat \imath} = -A_{- {\vec k}}^{\hat \imath} \equiv A_{\vec k}^{\hat \imath}$,
\begin{equation}
  A_{\vec k}^{\hat \imath}
  = i (v_{\vec k} \partial_{k_{\hat \imath}} v_{\vec k}^{\ast} - \partial_{k_{\hat \imath}} v_{\vec k} v_{\vec k}^{\ast})/2. \label{eq:sccon1}
\end{equation}
Correspondingly, the Berry curvatures change their signs; i.e,
$\Omega_{+ {\vec k} {\hat m}} = -\Omega_{- {\vec k} {\hat m}} \equiv \Omega_{{\vec k} {\hat m}}$,
\begin{equation}
  \Omega_{{\vec k} {\hat m}}
  = i \epsilon_{{\hat \imath} {\hat \jmath} {\hat m}} \partial_{k_{\hat \imath}} v_{\vec k} \partial_{k_{\hat \jmath}} v_{\vec k}^{\ast}. \label{eq:sccurv1}
\end{equation}
For chiral SCs with $w_{\vec k} \propto e^{i m \phi}$ in the polar coordinate ${\vec k} = (k \sin \theta \cos \phi, k \sin \theta \sin \phi, k \cos \theta)$,
the Berry connection and curvature are given by
\begin{subequations} \begin{align}
  A_{\vec k}^x
  = & -m |v_{\vec k}|^2 \sin \phi/k \sin \theta, \label{eq:sccon2x} \\
  A_{\vec k}^y
  = & m |v_{\vec k}|^2 \cos \phi/k \sin \theta, \label{eq:sccon2y} \\
  A_{\vec k}^z
  = & 0,  \label{eq:sccon2z}
\end{align} \label{eq:sccon2} \end{subequations}
and
\begin{subequations} \begin{align}
  \Omega_{{\vec k} z}
  = & m (k \partial_k + \cot \theta \partial_{\theta}) |v_{\vec k}|^2/k^2, \label{eq:sccurv2z} \\
  \Omega_{{\vec k} x}
  = & m \cos \phi (-k \cot \theta \partial_k + \partial_{\theta}) |v_{\vec k}|^2/k^2, \label{eq:sccurv2x} \\
  \Omega_{{\vec k} y}
  = & m \sin \phi (-k \cot \theta \partial_k + \partial_{\theta}) |v_{\vec k}|^2/k^2, \label{eq:sccurv2y}
\end{align} \label{eq:sccurv2} \end{subequations}
respectively.
Consequently, the AM for gapped chiral SCs at zero temperature is given by
\begin{equation}
  L_z
  = -\hbar \sum_{\vec k} ({\vec A}_{\vec k} \times {\vec k})_z
  = \hbar m \sum_{\vec k} |v_{\vec k}|^2
  = \hbar m N_0/2, \label{eq:scam}
\end{equation}
where we introduce the number of electrons at zero temperature $N_0 = \sum_{\vec k} 2 |v_{\vec k}|^2$.
This result is consistent with the recent microscopic studies suggesting $\gamma = 0$~\cite{Ishikawa01061977,Ishikawa01041980,PhysRevB.21.980,volovik1995,%
JPSJ.67.216,Goryo1998549,PhysRevB.69.184511,PhysRevB.84.214509,PhysRevB.85.100506}.
However, the bulk AM is well defined only in a gapped system at zero temperature.
Therefore, we cannot apply this result to three-dimensional chiral SCs with point or line nodes.
Below we calculate the HV, which is well defined in a gapless system or at finite temperature.

\subsection{Hall Viscosity in Two Dimensions} \label{sub:2d}
First, we consider $w_{\vec k} = e^{i \ell \phi}$ in two dimensions.
For odd $\ell$, the system is triplet and is classified into a class-D topological SC in terms of the topological periodic table~\cite{PhysRevB.78.195125,kitaev2009}.
On the other hand, for even $\ell$, it is singlet and is classified into a class-C topological SC.
Anyway it is gapped and the AM Eq.~\eqref{eq:scam} is well defined at zero temperature.
Since there is the rotational symmetry, only Eq.~\eqref{eq:etah} is nonzero,
\begin{align}
  2 \eta_{\rm H}
  = & -\frac{\hbar}{2} \sum_{\vec k} k^2 \Omega_{{\vec k} z} [1 - 2 f(E_{\vec k})] \notag \\
  = & -\frac{\hbar \ell}{2} \sum_{\vec k} k \partial_k |v_{\vec k}|^2 [1 - 2 f(E_{\vec k})]. \label{eq:sceta2d1}
\end{align}
At zero temperature, we employ the partial integral to obtain $2 \eta_{\rm H} = \hbar \ell N_0/2$,
which is consistent with the previous result for gapped chiral SCs~\cite{PhysRevB.79.045308,PhysRevB.84.085316,PhysRevB.86.245309}.

\begin{figure}[t]
  \centering
  \includegraphics[clip,width=0.48\textwidth]{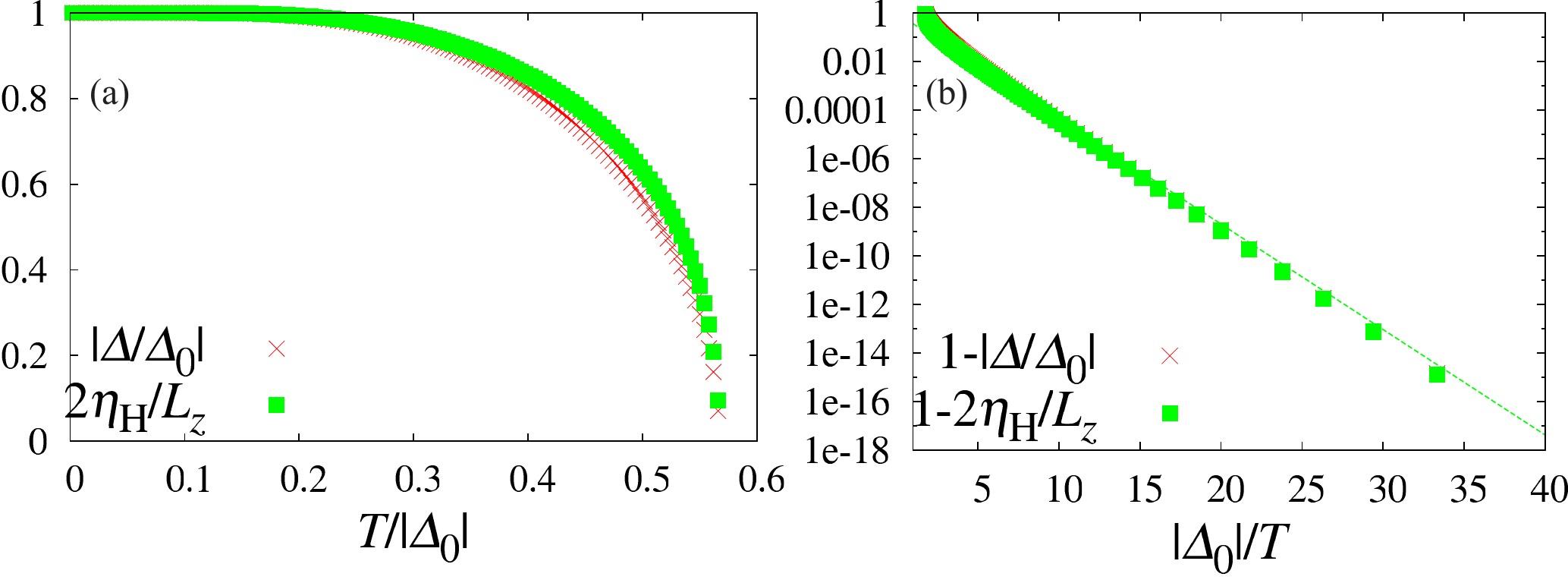}
  \caption{%
  (Color online) The gap strength (red cross) and the HV (green filled square) for two-dimensional chiral SCs as functions of
  (a) temperature $T/|\Delta_0|$ and (b) inverse temperature $|\Delta_0|/T$.
  The green broken line indicates $e^{-|\Delta_0|/T}$.%
  } \label{fig:2d}
\end{figure}

Let us turn to finite temperature.
By solving the gap equation Eq.~\eqref{eq:gap} and using the temperature dependence of the gap strength,
we evaluate the HV normalized by that at zero temperature, i.e., half the AM,
\begin{equation}
  \frac{2 \eta_{\rm H}}{\hbar \ell N_0/2}
  = \int_0^{\infty} d \xi \partial_{\xi} (\xi/E) [1 - 2 f(E)], \label{eq:sceta2d2}
\end{equation}
As shown in Fig.~\ref{fig:2d}, both the gap strength and the HV exponentially converge since the system is gapped.

\subsection{Hall Viscosity in Three Dimensions} \label{sub:3d}
Next, we consider $w_{\vec k} = \sqrt{4 \pi} Y_{\ell m}(\theta, \phi)$ in three dimensions,
where $Y_{\ell m}$ is a spherical harmonic function.
As emphasized above, three-dimensional chiral SCs have nodes generally, and the bulk AM is not well defined.
For example, $p_x \pm i p_y$ with $(\ell, m) = (1, \pm 1)$ and $d_{x^2 - y^2} \pm i d_{xy}$ with $(\ell, m) = (2, \pm 2)$ have point nodes at poles,
while $d_{zx} \pm i d_{yz}$ with $(\ell, m) = (2, \pm 1)$ has both point and line nodes.
Since there is the axial symmetry along the $z$ axis, two components Eq.~\eqref{eq:eta3d} are nonzero,
\begin{subequations} \begin{align}
  2 \eta^{(xx)(xy)}
  = & -\hbar \sum_{\vec k} k^{x 2} \Omega_{{\vec k} z} [1 - 2 f(E_{\vec k})] \notag \\
  = & -\hbar m \sum_{\vec k} \sin^2 \theta \cos^2 \phi (k \partial_k + \cot \theta \partial_{\theta}) |v_{\vec k}|^2 \notag \\
  & \times [1 - 2 f(E_{\vec k})], \label{eq:sceta1a} \\
  2 \eta^{(zx)(yz)}
  = & -\frac{\hbar}{2} \sum_{\vec k} (k^{z 2} \Omega_{{\vec k} z} - k^z k^x \Omega_{{\vec k} x} - k^z k^y \Omega_{{\vec k} y}) \notag \\
  & \times [1 - 2 f(E_{\vec k})] \notag \\
  = & -\frac{\hbar m}{2} \sum_{\vec k} (2 \cos^2 \theta k \partial_k + \cos 2 \theta \cot \theta \partial_{\theta}) |v_{\vec k}|^2 \notag \\
  & \times [1 - 2 f(E_{\vec k})], \label{eq:sceta1b}
\end{align} \label{eq:sceta1} \end{subequations}
both of which are reduced to $\hbar m N_0/2$ at zero temperature.

\begin{figure}[t]
  \centering
  \includegraphics[clip,width=0.48\textwidth]{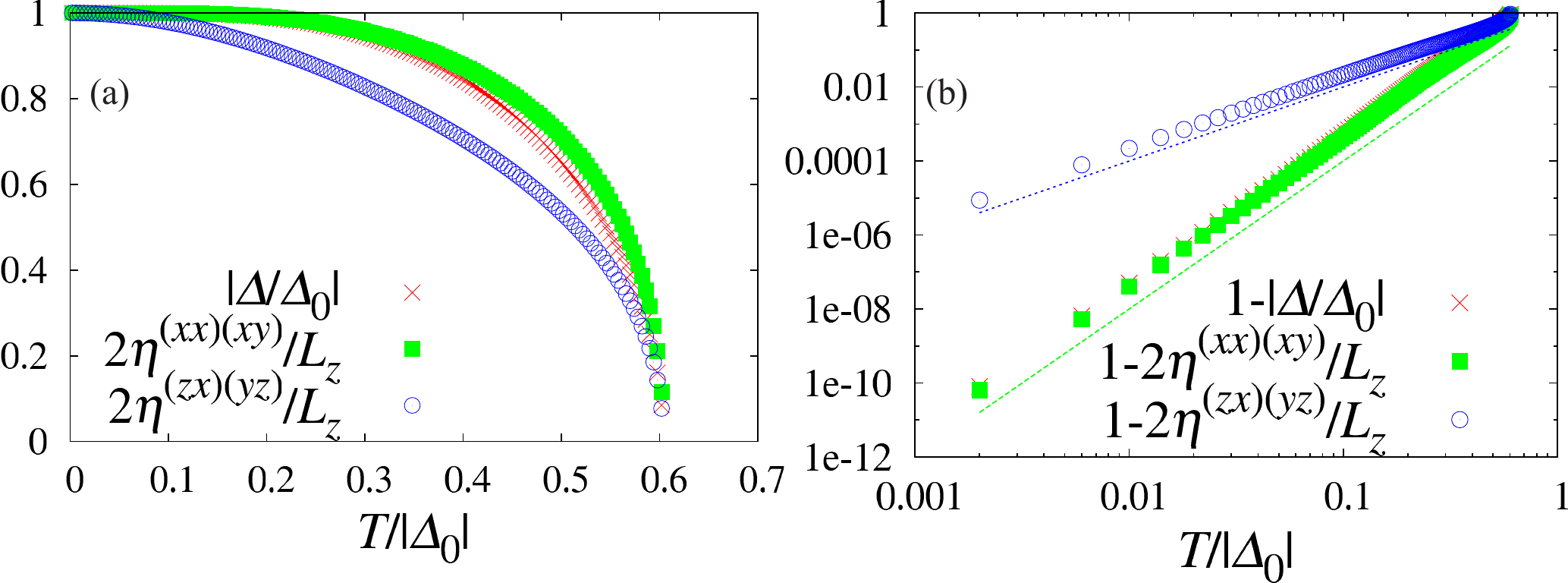}
  \caption{%
  (Color online) Temperature dependences of $|\Delta/\Delta_0|$ (red cross),
  $2 \eta^{(xx)(xy)}/(\hbar m N_0/2)$ (green filled square),
  and $2 \eta^{(zx)(yz)}/(\hbar m N_0/2)$ (blue open circle) for a three-dimensional $p_x + i p_y$ SC with $(\ell, m) = (1, 1)$.
  The green broken and blue dotted lines indicate $(T/|\Delta_0|)^4$ and $(T/|\Delta_0|)^2$, respectively.%
  } \label{fig:3d11}
\end{figure}

\begin{figure}[t]
  \centering
  \includegraphics[clip,width=0.48\textwidth]{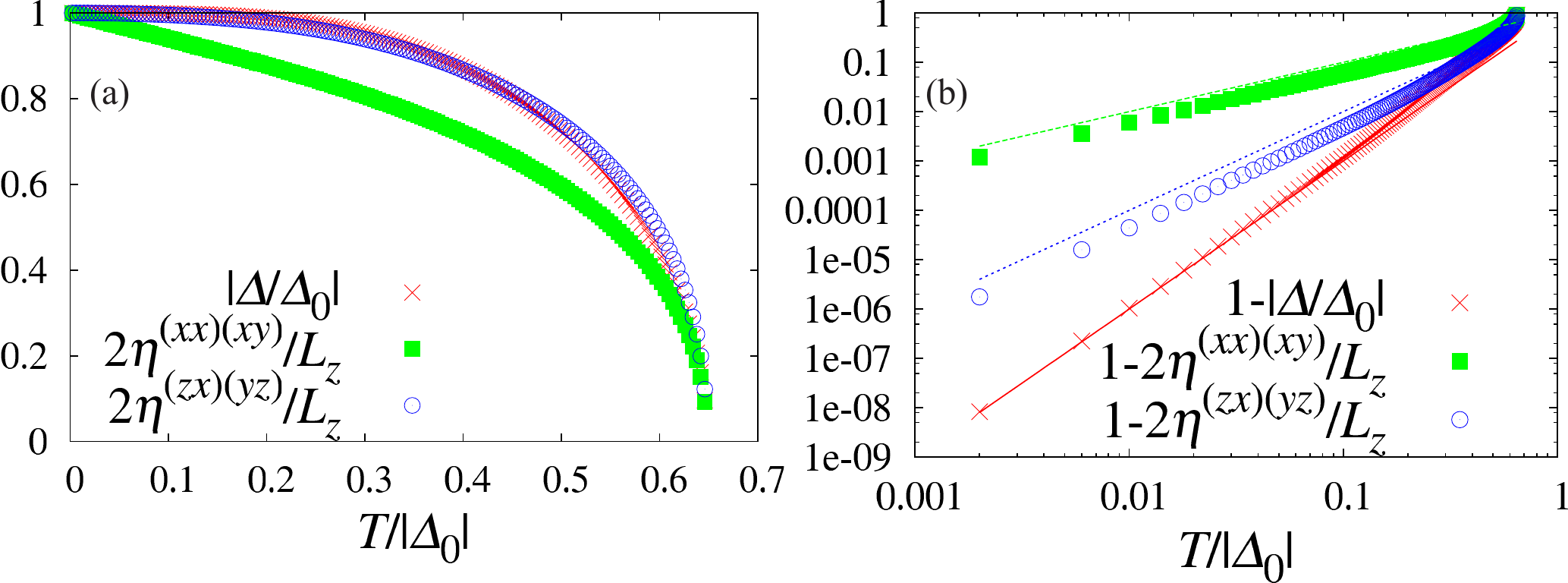}
  \caption{%
  (Color online) Temperature dependences of $|\Delta/\Delta_0|$ (red cross),
  $2 \eta^{(xx)(xy)}/(\hbar m N_0/2)$ (green filled square),
  and $2 \eta^{(zx)(yz)}/(\hbar m N_0/2)$ (blue open circle) for a three-dimensional $d_{zx} + i d_{yz}$ SC with $(\ell, m) = (2, 1)$.
  The red solid, green broken, and blue dotted lines indicate $(T/|\Delta_0|)^3$, $T/|\Delta_0|$, and $(T/|\Delta_0|)^2$, respectively.%
  } \label{fig:3d21}
\end{figure}

\begin{figure}[t]
  \centering
  \includegraphics[clip,width=0.48\textwidth]{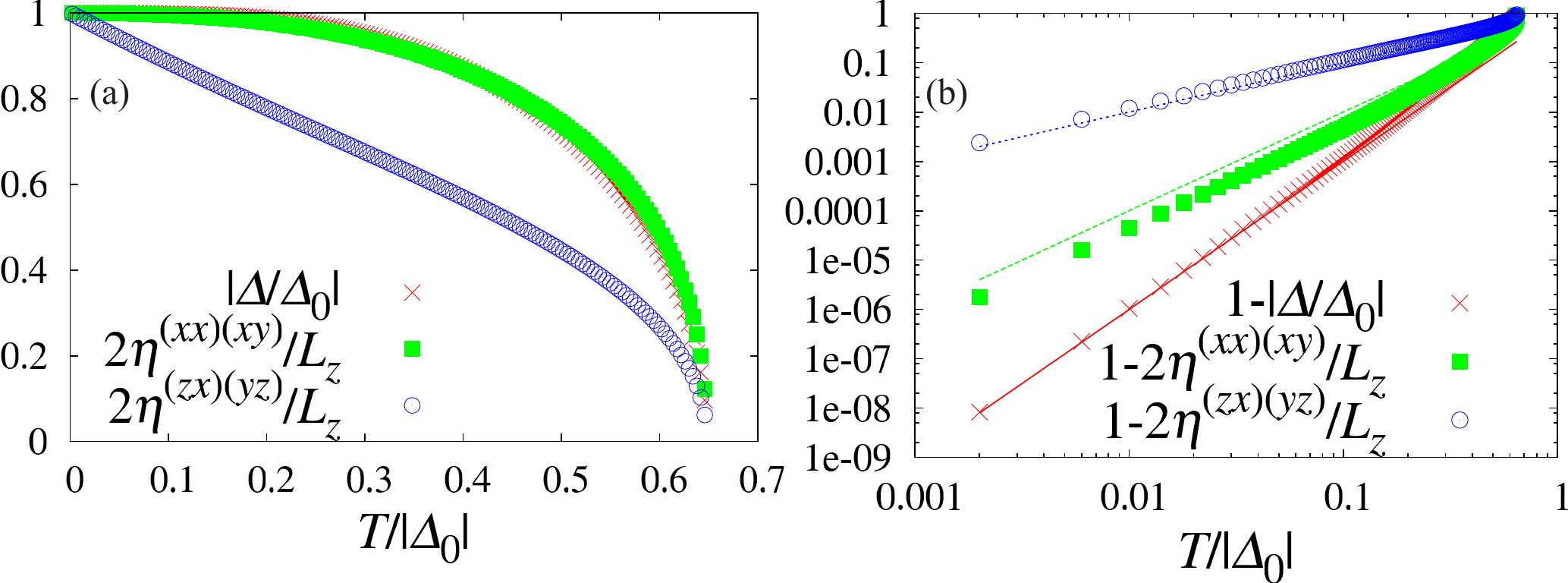}
  \caption{%
  (Color online) Temperature dependences of $|\Delta/\Delta_0|$ (red cross),
  $2 \eta^{(xx)(xy)}/(\hbar m N_0/2)$ (green filled square),
  and $2 \eta^{(zx)(yz)}/(\hbar m N_0/2)$ (blue open circle) for a three-dimensional $d_{x^2-y^2} + i d_{xy}$ SC with $(\ell, m) = (2, 2)$.
  The red solid, green broken, and blue dotted lines indicate $(T/|\Delta_0|)^3$, $(T/|\Delta_0|)^2$, and $T/|\Delta_0|$, respectively.%
  } \label{fig:3d22}
\end{figure}

We numerically calculate the normalized HV at finite temperature,
\begin{subequations} \begin{align}
  \frac{2 \eta^{(xx)(xy)}}{\hbar m N_0/2}
  = & \frac{3}{4} \int_0^{\infty} d \xi \int_0^{\pi} \sin \theta d \theta \sin^2 \theta \partial_{\xi} (\xi/E) [1 - 2 f(E)], \label{eq:sceta2a} \\
  \frac{2 \eta^{(zx)(yz)}}{\hbar m N_0/2}
  = & \frac{3}{2} \int_0^{\infty} d \xi \int_0^{\pi} \sin \theta d \theta \cos^2 \theta \partial_{\xi} (\xi/E) [1 - 2 f(E)], \label{eq:sceta2b}
\end{align} \label{eq:sceta2} \end{subequations}
instead of Eq.~\eqref{eq:sceta1}.
In Figs.~\ref{fig:3d11}, \ref{fig:3d21}, and \ref{fig:3d22},
we show the temperature dependences of the gap strength and the HV for $p_x + i p_y$, $d_{zx} + i d_{yz}$, and $d_{x^2-y^2} + i d_{xy}$, respectively.
All of them converge by the power laws since the systems have nodes.
These powers can be analytically obtained by expanding the integrands in Eqs.~\eqref{eq:gap} and \eqref{eq:sceta2} around nodes as in the appendix.
Especially for $d_{x^2 - y^2} + i d_{xy}$, we find that
the temperature dependence of $|\Delta/\Delta_0|$ and $2 \eta^{(xx)(xy)}/(\hbar m N_0/2)$ is almost determined by  the line node,
while that of $2 \eta^{(zx)(yz)}/(\hbar m N_0/2)$ is by the point nodes.
We also summarize their powers in Table ~\ref{tab:pow}.
\begin{table}[t]
  \centering
  \caption{%
  Power-law behaviors of the gap strength and the HV at low temperature for several pairing symmetries in three dimensions.
  See also Figs.~\ref{fig:3d11}, \ref{fig:3d21}, and \ref{fig:3d22}.%
  } \label{tab:pow}
  \begin{tabular}{cccccc} \hline \hline
    symmetry & $(\ell, m)$ & nodes & $|\Delta|$ & $2 \eta^{(xx)(xy)}$ & $2 \eta^{(zx)(yz)}$ \\ \hline
    $p_x + i p_y$ & $(1, 1)$ & point & $T^4$ & $T^4$ & $T^2$ \\
    $d_{zx} + i d_{yz}$ & $(2, 1)$ & point and line & $T^3$ & $T$ & $T^2$ \\
    $d_{x^2-y^2} + i d_{xy}$ & $(2, 2)$ & point & $T^3$ & $T^2$ & $T$ \\ \hline \hline
  \end{tabular}
\end{table}

\section{Discussion and Summary} \label{sec:summary}
The temperature dependence of the AM for $p_x + i p_y$ was calculated in finite systems such as
a mesoscopic cylinder~\cite{JPSJ.67.216} and a macroscopic disk~\cite{PhysRevB.84.214509,PhysRevB.85.100506} compared to the coherence length.
In three dimensions, the AM decreases from $\hbar N_0/2$ by $T^2$ owing to the presence of the point nodes~\cite{JPSJ.67.216}.
In two dimensions, the AM also decreases from $\hbar N_0/2$ by $T^2$,
which is attributed to the presence of the Majorana edge modes~\cite{PhysRevB.84.214509,PhysRevB.85.100506}.
Although the HV is equal to half the AM at zero temperature not only in two dimensions but also in three dimensions,
the temperature dependence of the AM and the HV is generally different.
This discrepancy does not conflict with the relation between the AM and the HV~\cite{PhysRevB.79.045308,PhysRevB.84.085316},
because it is available only for gapped systems at zero temperature.
Indeed such a difference was recently found in two-dimensional gapless systems by using the holographic approach~\cite{1403.6047}, too.
As pointed out in this paper, the AM is given by the polarization of the momentum.
In this sense, it is seen as a thermodynamic quantity, while the HV is just a transport coefficient but not a bulk quantity.
Therefore it is natural to think that physical origins of those quantities and also their temperature dependences are different in general,
even if they have a special relation at zero temperature.
It may be interesting to study other models, e.g., quantum Hall systems and chiral SCs with higher $\ell$, at finite temperature.
Let us note that such a relation between a thermodynamic quantity and a transport coefficient
was also proposed for the entropy density and the shear viscosity~\cite{PhysRevLett.94.111601}.
Nevertheless it is expected that they would have a universal relation only in the extremely strong coupling situation.

To summarize, we derive the Berry-phase formulas for the AM and the HV by using the Keldysh formalism in a curved spacetime.
First, we examine the physical quantity conjugate to the AM, namely, an angular velocity of rotation, from the gauge-theoretical viewpoint of gravity.
Since an angular velocity is the antisymmetric torsional electric field,
we define the AM by the temporal integral of the antisymmetric momentum current induced by an adiabatic deformation, which is implemented by the gradient expansion.
Viscosity is the response of the stress tensor, i.e., the symmetric momentum current, to a strain-rate tensor.
It can be derived by the perturbation theory with respect to torsion because a strain-rate tensor is described by the symmetric torsional electric field.
We also apply these results to chiral SCs in two and three dimensions.
In two dimensions at zero temperature, we reproduce $L_z = 2 \eta_{\rm H} = \hbar \ell N_0/2$ without any finite-size effects.
In three dimensions at zero temperature, where the AM is not well defined owing to the presence of nodes,
we find that the HV is equal to half the AM calculated in finite systems previously.
Although it is not related to the AM at finite temperature, it is useful to determine the gap structure for chiral SCs.

\begin{acknowledgments}
  We thank H. Sumiyoshi and Y. Tada for fruitful discussions.
  This work was supported by Grants-in-Aid for the Japan Society for the Promotion of Science, Fellows No.~$24$-$600$ and No.~$25$-$4302$.
\end{acknowledgments}

\appendix
\section{Temperature Dependence of Hall Viscosity} \label{sec:temp}
In this appendix, we approximately but analytically calculate the gap strength and the HV to explain their power-law behaviors at low temperature.
\begin{widetext}
For $p_x + i p_y$, where $|w_{\vec k}|^2 = 3 \sin^2 \theta/2$,
we expand Eqs.~\eqref{eq:gap} and \eqref{eq:sceta2} around the point node $\theta = 0$ to obtain
\begin{subequations} \begin{align}
  \ln |\Delta_0/\Delta|
  \simeq & 2 \int_0^{\infty}d \xi \int_0^{\pi/2} \theta d \theta
  \frac{3 \theta^2/2}{\sqrt{\xi^2 + 3 |\Delta|^2 \theta^2/2}} (e^{\beta \sqrt{\xi^2 + 3 |\Delta|^2 \theta^2/2}} + 1)^{-1} \notag \\
  = & \frac{4}{3} (T/|\Delta|)^4 \int_0^{\infty} d x \int_0^{\sqrt{3/2} \beta |\Delta| \pi/2} d y \frac{y^3}{r} (e^r + 1)^{-1}
  \to \frac{7 \pi^4}{135} (T/|\Delta_0|)^4, \label{eq:gap3d11} \\
  1 - \frac{2 \eta^{(xx)(xy)}}{\hbar m N_0/2}
  \simeq & 3 \int_0^{\infty} d \xi \int_0^{\pi/2} \theta d \theta \theta^2
  \frac{3 |\Delta|^2 \theta^2/2}{(\xi^2 + 3 |\Delta|^2 \theta^2/2)^{3/2}} (e^{\beta \sqrt{\xi^2 + 3 |\Delta|^2 \theta^2/2}} + 1)^{-1} \notag \\
  = & \frac{4}{3} (T/|\Delta|)^4 \int_0^{\infty} d x \int_0^{\sqrt{3/2} \beta |\Delta| \pi/2} d y \frac{y^5}{r^3} (e^r + 1)^{-1}
  \to \frac{28 \pi^4}{675} (T/|\Delta_0|)^4, \label{eq:sceta3d11a} \\
  1 - \frac{2 \eta^{(zx)(yz)}}{\hbar m N_0/2}
  \simeq & 6 \int_0^{\infty} d \xi \int_0^{\pi/2} \theta d \theta
  \frac{3 |\Delta|^2 \theta^2/2}{(\xi^2 + 3 |\Delta|^2 \theta^2/2)^{3/2}} (e^{\beta \sqrt{\xi^2 + 3 |\Delta|^2 \theta^2/2}} + 1)^{-1} \notag \\
  = & 4 (T/|\Delta|)^2 \int_0^{\infty} d x \int_0^{\sqrt{3/2} \beta |\Delta| \pi/2} d y \frac{y^3}{r^3} (e^r + 1)^{-1}
  \to \frac{2 \pi^2}{9} (T/|\Delta_0|)^2. \label{eq:sceta3d11b}  
\end{align} \label{eq:3d11} \end{subequations}
In the second lines, we change the variables by $x = \beta \xi$ and $y = \sqrt{3/2} \beta |\Delta| \theta$,
and in the third lines, we take the low-temperature limit $\sqrt{3/2} \beta |\Delta| \pi/2 \to \infty$.
These integrals can be analytically estimated in the polar coordinate.

For $d_{zx} + i d_{yz}$, where $|w_{\vec k}|^2 = 15 \sin^2 \theta \cos^2 \theta/2$, there are both point and line nodes.
First, we expand Eqs.~\eqref{eq:gap} and \eqref{eq:sceta2} around the point node $\theta = 0$
and change the variables by $x = \beta \xi$ and $y = \sqrt{15/2} \beta |\Delta| \theta$, which results in
\begin{subequations} \begin{align}
  \ln |\Delta_0/\Delta|
  \simeq & 2 \int_0^{\infty}d \xi \int_0^{\pi/2} \theta d \theta
  \frac{15 \theta^2/2}{\sqrt{\xi^2 + 15 |\Delta|^2 \theta^2/2}} (e^{\beta \sqrt{\xi^2 + 15 |\Delta|^2 \theta^2/2}} + 1)^{-1} \notag \\
  = & \frac{4}{15} (T/|\Delta|)^4 \int_0^{\infty} d x \int_0^{\sqrt{15/2} \beta |\Delta| \pi/2} d y \frac{y^3}{r} (e^r + 1)^{-1}
  \to \frac{7 \pi^4}{675} (T/|\Delta_0|)^4, \label{eq:gap3d21-p} \\
  1 - \frac{2 \eta^{(xx)(xy)}}{\hbar m N_0/2}
  \simeq & 3 \int_0^{\infty} d \xi \int_0^{\pi/2} \theta d \theta \theta^2
  \frac{15 |\Delta|^2 \theta^2/2}{(\xi^2 + 15 |\Delta|^2 \theta^2/2)^{3/2}} (e^{\beta \sqrt{\xi^2 + 15 |\Delta|^2 \theta^2/2}} + 1)^{-1} \notag \\
  = & \frac{4}{75} (T/|\Delta|)^4 \int_0^{\infty} d x \int_0^{\sqrt{15/2} \beta |\Delta| \pi/2} d y \frac{y^5}{r^3} (e^r + 1)^{-1}
  \to \frac{28 \pi^4}{16875} (T/|\Delta_0|)^4, \label{eq:sceta3d21a-p} \\
  1 - \frac{2 \eta^{(zx)(yz)}}{\hbar m N_0/2}
  \simeq & 6 \int_0^{\infty} d \xi \int_0^{\pi/2} \theta d \theta
  \frac{15 |\Delta|^2 \theta^2/2}{(\xi^2 + 15 |\Delta|^2 \theta^2/2)^{3/2}} (e^{\beta \sqrt{\xi^2 + 15 |\Delta|^2 \theta^2/2}} + 1)^{-1} \notag \\
  = & \frac{4}{5} (T/|\Delta|)^2 \int_0^{\infty} d x \int_0^{\sqrt{15/2} \beta |\Delta| \pi/2} d y \frac{y^3}{r^3} (e^r + 1)^{-1}
  \to \frac{2 \pi^2}{45} (T/|\Delta_0|)^2. \label{eq:sceta3d21b-p}  
\end{align} \label{eq:3d21-p} \end{subequations}
On the other hand, for the line node, we redefine $\theta \to \pi/2 - \theta$ and expand Eqs.~\eqref{eq:gap} and \eqref{eq:sceta2} around $\theta = 0$, leading to
\begin{subequations} \begin{align}
  \ln |\Delta_0/\Delta|
  \simeq & 2 \int_0^{\infty}d \xi \int_0^{\pi/2} d \theta
  \frac{15 \theta^2/2}{\sqrt{\xi^2 + 15 |\Delta|^2 \theta^2/2}} (e^{\beta \sqrt{\xi^2 + 15 |\Delta|^2 \theta^2/2}} + 1)^{-1} \notag \\
  = & 2 \sqrt{\frac{2}{15}} (T/|\Delta|)^3 \int_0^{\infty} d x \int_0^{\sqrt{15/2} \beta |\Delta| \pi/2} d y \frac{y^2}{r} (e^r + 1)^{-1}
  \to \frac{\sqrt{3} \pi \zeta(3)}{2 \sqrt{10}} (T/|\Delta_0|)^3, \label{eq:gap3d21-l} \\
  1 - \frac{2 \eta^{(xx)(xy)}}{\hbar m N_0/2}
  \simeq & 3 \int_0^{\infty} d \xi \int_0^{\pi/2} d \theta
  \frac{15 |\Delta|^2 \theta^2/2}{(\xi^2 + 15 |\Delta|^2 \theta^2/2)^{3/2}} (e^{\beta \sqrt{\xi^2 + 15 |\Delta|^2 \theta^2/2}} + 1)^{-1} \notag \\
  = & 3 \sqrt{\frac{2}{15}} (T/|\Delta|) \int_0^{\infty} d x \int_0^{\sqrt{15/2} \beta |\Delta| \pi/2} d y \frac{y^2}{r^3} (e^r + 1)^{-1}
  \to \frac{\sqrt{3} \pi \ln 2}{2 \sqrt{10}} (T/|\Delta_0|), \label{eq:sceta3d21a-l} \\
  1 - \frac{2 \eta^{(zx)(yz)}}{\hbar m N_0/2}
  \simeq & 6 \int_0^{\infty} d \xi \int_0^{\pi/2} d \theta \theta^2
  \frac{15 |\Delta|^2 \theta^2/2}{(\xi^2 + 15 |\Delta|^2 \theta^2/2)^{3/2}} (e^{\beta \sqrt{\xi^2 + 15 |\Delta|^2 \theta^2/2}} + 1)^{-1} \notag \\
  = & \frac{4 \sqrt{2}}{5 \sqrt{15}} (T/|\Delta|)^3 \int_0^{\infty} d x \int_0^{\sqrt{15/2} \beta |\Delta| \pi/2} d y \frac{y^4}{r^3} (e^r + 1)^{-1}
  \to \frac{3 \sqrt{3} \pi \zeta(3)}{20 \sqrt{10}} (T/|\Delta_0|)^3. \label{eq:sceta3d21b-l}  
\end{align} \label{eq:3d21-l} \end{subequations}
By comparing each power,
the gap strength $|\Delta/\Delta_0|$ and one component of the HV $2 \eta^{(xx)(xy)}/(\hbar m N_0/2)$ are mainly contributed from the line node,
while the other component $2 \eta^{(zx)(yz)}/(\hbar m N_0/2)$ is from the point nodes.

For $d_{x^2 - y^2} + i d_{xy}$, where $|w_{\vec k}|^2 = 15 \sin^4 \theta/8$,
we expand Eqs.~\eqref{eq:gap} and \eqref{eq:sceta2} around the point node $\theta = 0$ and introduce $y = \sqrt{15/8} \beta |\Delta| \theta^2$.
Then we obtain
\begin{subequations} \begin{align}
  \ln |\Delta_0/\Delta|
  \simeq & 2 \int_0^{\infty}d \xi \int_0^{\pi/2} \theta d \theta
  \frac{15 \theta^4/8}{\sqrt{\xi^2 + 15 |\Delta|^2 \theta^4/8}} (e^{\beta \sqrt{\xi^2 + 15 |\Delta|^2 \theta^4/8}} + 1)^{-1} \notag \\
  = & \sqrt{\frac{8}{15}} (T/|\Delta|)^3 \int_0^{\infty} d x \int_0^{\sqrt{15/8} \beta |\Delta| (\pi/2)^2} d y \frac{y^2}{r} (e^r + 1)^{-1}
  \to \frac{\sqrt{3} \pi \zeta(3)}{2 \sqrt{10}} (T/|\Delta_0|)^3, \label{eq:gap3d22} \\
  1 - \frac{2 \eta^{(xx)(xy)}}{\hbar m N_0/2}
  \simeq & 3 \int_0^{\infty} d \xi \int_0^{\pi/2} \theta d \theta \theta^2
  \frac{15 |\Delta|^2 \theta^4/8}{(\xi^2 + 15 |\Delta|^2 \theta^4/8)^{3/2}} (e^{\beta \sqrt{\xi^2 + 15 |\Delta|^2 \theta^4/8}} + 1)^{-1} \notag \\
  = & \frac{4}{5} (T/|\Delta|)^2 \int_0^{\infty} d x \int_0^{\sqrt{15/8} \beta |\Delta| (\pi/2)^2} d y \frac{y^3}{r^3} (e^r + 1)^{-1}
  \to \frac{2 \pi^2}{45} (T/|\Delta_0|)^2, \label{eq:sceta3d22a} \\
  1 - \frac{2 \eta^{(zx)(yz)}}{\hbar m N_0/2}
  \simeq & 6 \int_0^{\infty} d \xi \int_0^{\pi/2} \theta d \theta
  \frac{15 |\Delta|^2 \theta^4/8}{(\xi^2 + 15 |\Delta|^2 \theta^4/8)^{3/2}} (e^{\beta \sqrt{\xi^2 + 15 |\Delta|^2 \theta^4/8}} + 1)^{-1} \notag \\
  = & 3 \sqrt{\frac{8}{15}} (T/|\Delta|) \int_0^{\infty} d x \int_0^{\sqrt{15/8} \beta |\Delta| (\pi/2)^2} d y \frac{y^2}{r^3} (e^r + 1)^{-1}
  \to \frac{\sqrt{3} \pi \ln 2}{\sqrt{10}} (T/|\Delta_0|). \label{eq:sceta3d22b}
\end{align} \label{eq:3d22} \end{subequations}
Thus, all the power-law behaviors can be explained by nodal excitations.
\end{widetext}
%
\end{document}